\documentstyle[12pt]{article}
\title  {Scalar products of elementary distributions}

\author { Philippe Droz-Vincent\\[2mm]LUTH\\
Meudon \footnote{Observatoire de Paris, CNRS, Universit\'e Paris Diderot,
5  place Jules Janssen,   92195  Meudon, France   }}

\date { \      }

\newcommand {\batonR}{ \mbox{I\hspace*{-1mm}R} }
\newcommand {\batonL}{ \mbox{I\hspace*{-1mm}L} }
\newcommand {\batonK}{ \mbox{I\hspace*{-1mm}K} }
\newcommand {\batonP}{ \mbox{I\hspace*{-1mm}P} }
\newcommand {\batonF}{ \mbox{I\hspace*{-1mm}F} }
\newcommand {\batonB}{ \mbox{I\hspace*{-1mm}B} }
\newcommand {\batonC}{ \mbox{I\hspace*{-2mm}C} }
\newcommand {\avragzer}{{\scriptscriptstyle < 0 > }}
\newcommand {\zero}{ {<0>  } }

\newcommand {\plu}{ {\scriptscriptstyle  +}}
\newcommand {\moi}{ {\scriptscriptstyle  -}}
\newcommand {\Del}{\Delta}

\newcommand {\calS}{  {\cal S}  }
\newcommand {\calA}{  {\cal A}  }
\newcommand{\calp}{ {\cal P} }
\newcommand{\calh}{ {\cal H} }
\newcommand{\calj}{ {\cal  J }}
\newcommand{\calB}{ {\cal B} }

\newcommand{\eron}{ {\cal E} }
\newcommand{\cpx}{ {\scriptstyle C}  }

\newcommand{\beq}{\begin{equation}}
\newcommand{\eeq}{\end{equation}}
\newcommand{\lam}{\lambda}

 \newcommand {\half}{ {1 \over 2}}
 \newcommand {\noi}{\noindent}
\newcommand {\qu}{{(q)}}
 \newcommand{\degr}{{\rm deg}}
\newcommand {\eme}{ {(m)}   }
\newcommand{\ere}{ {(r)} }
\newcommand {\ene}{ {(n)}  }
\newcommand {\pe}{ {(p)} }

\newcommand {\del}{\delta}
   
\newcommand{\Plex}{{\mbox{I\hspace*{-1mm}P}}^{\scriptstyle C}}
\newcommand{\Flex}{{\mbox{I\hspace*{-1mm}F}}^{\scriptstyle C}}
\newcommand {\Darp}{\Delta ^\sharp}
\newcommand {\alp}{\alpha}

\newcommand{\calq}{ {\cal Q} }
\newcommand{\calw}{ {\cal W} }
\newcommand  {\disp}{\displaystyle}
\newcommand{\cinf}{C_\infty}
 \newcommand {\cinfsq}{C^2 _\infty}
\newcommand {\brok}{ {\widehat C} ^2 _\infty}
  \newtheorem{prop}{Proposition}
 \newtheorem{theo}{Theorem}

\newcommand{\beprop}{\begin{prop}}
\newcommand{\betheo}{\begin{theo}}
\newcommand{\enprop}{\end{prop}}
\newcommand{\entheo}{\end{theo}}
\begin{document}
\maketitle
\abstract{The  field of real  numbers being  extended  as  a
larger   commutative  field, we investigate  the possibility  of
defining a  scalar product  for  the  distributions of finite discrete support.
Then  we  focus  on  the  most simple possible extension (which is  an ordered field), 
  we  provide   explicit  formulas for   this   scalar   product,  and  we   exhibit a structure of   positive
definite  inner-product space.
In a one dimensional application  to   the   Schroedinger  equation,     the  distributions   supported
  by  the origin   are embedded into a   bra-ket  vector space, where the "singular" potential
describing   point interaction is defined in a natural way.  
A contact with the hyperreal numbers  that arise in nonstandard analysis is possible
but not essential,   our  extensions of $\batonR$  and $\batonC$ being   obtained
 by  a  quite  elementary method. }

$$    \              $$

 \bigskip

\section {Introduction, Notation}

Many  difficulties of quantum mechanics stem from the need to
consider unnormalizable state vectors, which implies  going out of
a strict Hilbert space formulation. To some extent, using Dirac's
formalism  permits to ignore this complication  and all physicists
are familiar with wave "functions" that  can actually be distributions,
but this  approach runs into the problem of their norm.

\noi
 A more rigorous formulation resorts to the concept of  "rigged Hilbert  space"
\cite{gel} which considers generalized eigenvectors; however these
 more general state vectors still have a divergent norm in the usual sense.

\noi  The difficulty has two sources:

\noi  a) The usual scalar product is defined through the integral of   a
  product of  two distributions, and this product may be ill-defined.

\noi  b) Even if the multiplication of these distributions is well-defined, the
integral may be divergent.

In view of  case a) valuable work has been performed by mathematicians in order to
define the product of singular distributions.
 But the obligation to circumvent the well-known Schwartz's impossibility
theorem results in several complications that
are discouraging for the physicist. Some authors propose a  commutative product which  cannot
 be always associative~\cite{Li}.
 Colombeau  and followers  \cite{colomb}  succeeded in constructing a
 differential  algebra which is associative, but at the price of including new objects that
are not always distributions.

\noi  Nevertheless,   we can make the following observation:
Leaving apart some specific problems of field theory, what is most basically
 needed for quantum mechanics is a {\em scalar product} of distributions,
 rather than a multiplication which would produce another distribution.


\noi  According to this remark, in this paper we radically avoid to consider
the multiplication of distributions, and shall directly resort to their
{\em scalar product}.

\noi  But  it seems that no consistent picture can be obtained within the
 framework of real (or complex) numbers. Indeed intuition suggests
that some "infinite quantity" should be introduced. This situation
naturally leads to  an  enlargement of the field of real
(or complex) numbers, by embedding this field into some suitable
commutative and associative algebra (if possible a field) of which
some element somehow represents this "infinity". Naturally we
expect that this new algebra of scalars is of    infinite dimension
over the usual scalars.

\noi At a more elementary level,  we made an attempt some  years ago
\cite{LMP}
  defining a scalar product for distributions  of which the support is a
finite number of  isolated points.
This naive approach already allowed  for considering a space
of states endowed with  a sesquilinear   form (which actually  was
a Hermitian form).
But  quantum mechanics requires that at least  some  states  have  a
 {\em positive definite} norm,   whereas the algebra
of scalars considered in this early work  was  completely lacking of any
ordering relation (except in its trivial restriction to the reals).

\noi Therefore,  in   the present work  we aim  at    building a  new scalar
 product  which takes  on values  in  the  complexified of a  totally   ordered  field.
For the sake of  physical applications it  is  obviously desirable  that  this new
scalar product, as much as possible,  mimicks several   nice properties
 of the inner  product in Hilbert spaces.

\noi    Our main tool  is the  introduction of  new quantities that can be
either   infinitely large or  infinitesimal.
  Such an extension was advocated as soon as in 1975 by
  M.O.Farrukh~\cite{far}
who proposed to formulate quantum mechanics in  a {\em non-standard}  Hilbert space.
According to his approach, the divergent lenghts of unnormalizable states are
re-defined as  infinitely large non-standard   numbers.
In his framework however, the status of distributions suffers from some
complications (for instance, all representations of distributions
 as   pointwise-defined   non-standard functions
 coincide {\em except} on an infinitesimal neighborhood of zero).
In the same spirit,  see a recent work by  Almeida and Teixeira~\cite{alm}.

But all the works carried out along this line assume  the knowlege of nonstandard (n.s.)
analysis~\cite{rob}  which in turn  requires mastering mathematical logics, the theory of ultra filters,
and the Transfer Principle! In general it can be observed that the theory of   nonstandard  Hilbert spaces somehow  avoids the
 concept of distribution,  to a large extent replaced by that of  pointwise defined,   nonstandard-valued,  function  of a  n.s. variable.

  In contradistinction,  the  generalized numbers involved in  the present work  are constructed by   elementary methods, by-passing  all the  sophistications of  mathematical logics.
Moreover we restrict ourselves to functions  and distributions that depend on a real (and standard)  variable,  though they can be linearly combined  with  help of  coefficients that are  generalized numbers.
 In this framework we naturally keep treating the distributions on their own right,  and
 consider  differentiation {\em in the sense of distributions}.


\bigskip
In Section  2     we   recall some elementary results about 
 {\em bra-ket  vector  spaces} in general and, without specifying the field of scalars, we consider
the possibility of constructing  a Hermitian form $<.,.>$ on the space of
  the distributions  concentrated at the origin.

\noi 
In Section  3   we focus on the field $\batonF ^ C$ (which is in
some sense the most simple  extension of  $\batonC$) and construct  a positive definite
 Hermitian form $(.,.)$ on the space of polynomials.
We give explicit formulas for use in calculations and we generalize  $(.,.)$   to 
 more general regular functions that
are the  Fourier transformed of the  distributions with finite discrete support.
This situation feeds us back with a Hermitian form  $<.,.>$ defined on  the
 space of these distributions.

\noi Finally we display in Section 4 an application to  the  Schroedinger equation in one dimension with point interaction,  and  
 we construct  the solutions.

\subsection{Terminology}

\noi The words "scalar product"  have been loosely employed in the
literature,  including in previous works of the  author. We need
here a more precise  terminology~\cite{bog}.

\noi
Let   $   \batonL    \subset  \batonK  $  be   commutative fields with an involution
$*$,   such  that   $\batonK$  is a field extension  of  $\batonL$.
Let   $\calA  ,  \calB$    be    vector spaces on      $\batonL$.
The binary  {\sl  map}
$$  <.,.> :    \qquad        \calA  \times  \calB         \mapsto      \batonK              $$
that is  $<a,b>  \in  \batonK$     when    $a  \in  \calA$  and  $b  \in  \calB$,
 is   a   {\sl sesquilinear   map} when it is antilinear in $a$     and linear in $b$.

\noi  When $\calA =  \calB$  it may happen that  in addition we have
  $  <a,b> ^*  =  <b, a >$
for all couple $a,b$. In this case $<.,.>$ is a  {\sl  Hermitian map}:
$ \calA  \times  \calA    \mapsto  \batonK$.

\noi When further we have  $\batonK = \batonL$, this  map  is actually a
{\sl  Hermitian form}  on  $\calA$ and we say that $\calA$ is a
{\sl   bra-ket  vector space}.
In some sense a  bra-ket  vector space can be seen as an inner-product 
space which may be degenerated.  However, in agreement with a widespread but not universal convention \cite{bog} \cite{aziz} we reserve   the name inner-product space for  non-degenerate cases, as follows:

\noi  We say that $\cal A$ endowed with the Hermitian  form  $<.,.>$ is an
{\sl inner-product  space in the weak sense}
when

\beq <u,v> = 0  \    \quad    \forall v   \qquad  \quad      {\rm implies}
\quad
  \    u=0                                       \label{weak}  \eeq
We say that    $\cal A$ endowed with the Hermitian form is an {\sl
inner-product  space in the strong   sense}
when   there is no neutral  vector, in other words
\beq
<u , u> = 0 \      \qquad  \quad  {\rm implies} \      \
 u=0                 \label{strong}     \eeq


\noi  It is easy to check that condition  (\ref{strong}) implies
(\ref{weak}).
Indeed, if there exists $u$ such that  $ <u , v > = 0  \     \forall v$, in
particular   $  <u , u> $  vanishes, thus $u$ itself must be zero,
according to  (\ref{strong}). The converse is not true.

   It is more interesting to distinguish between the weak and
the strong definitions when   $\batonK$ is  the complexified of an {\em ordered extension}
$\batonB$  of $\batonR$,  say   $\batonK  =  \batonB ^C$ with an obvious notation  
 (we assume that the ordering is compatible with addition and multiplication).

\noi   In  this case, the  weak definition refers to a space with {\sl indefinite
metric}, because  it allows for positive, negative or vanishing
values of $ <u,u> $.

\noi   In contrast the strong definition refers to a
space with {\sl definite metric}; 
   in this case, the space $\cal A$  endowed with the form  $<.,.>$  is 
a {\sl   Hermitian    vector   space}.


\medskip
\noi
The following result is straightforward and will be  useful  in the sequel,
\beprop
Let  $\calh _1$ and $\calh _2 $ be {\em   bra-ket  } vector  spaces 
( on $\batonL  \subset  \batonK$), 
 with  $\batonK$-valued  Hermitian forms
$<.,.> _1  $  and  $<.,.> _2 $ respectively,   and such that
$\calh _1 \cap  \calh _2  = \{  0  \} $.
  Defining  further two sesquilinear   {\em maps}
$$  \ll.,.> :  \qquad    \calh _1  \times  \calh _2   \mapsto      {\em  \batonK }  $$
$$  <.,\gg :   \qquad     \calh _2  \times  \calh _1   \mapsto   {\em  \batonK }     $$
amounts to construct   a sesquilinear  map:
             $        \calh _1     \oplus  \calh _2       \mapsto       \batonK    $.
When, in addition, we  have  that
$\ll  f_1 , f_2>  ^*   = <f_2 , f_1\gg  $    for all   $f_1 \in  \calh_1   ,  f_2 \in \calh _2 $,
then this sesquilinear  map      actually   is    a   {\em  Hermitian }  map,
 and  if    $\batonL  =  \batonK$,   we  get  on        $ \calh _1     \oplus  \calh _2   $
 a    Hermitian  {\em form }       which 
 encompasses   $<.,.> _1  $  and  $<.,.> _2 $.

\noi
The  direct sum  {\em  is   not  orthogonal}  unless     $ \ll  ,.>   $    and    $   <.,.\gg $
both   vanish.
\enprop

\beprop
Let  $\eron _ 1$ and $\eron _ 2$ be  {\em  inner-product spaces } such that
$\eron _1 \cap  \eron _2  = \{  0  \} $. There is a unique  {\em orthogonal} direct sum
$\eron _ 1 \oplus  \eron _ 2 $ and this direct sum is an inner-product space.
\enprop
 In the situation considered here, it is essential that  $\eron _1 $  and  $ \eron  _2$
 are  mutually orthogonal.  Otherwize, there is no unicity and the last  statement might
 be wrong.

\bigskip
\noi {\bf Extension of the scalars}

\noi   In the course of this   article we  generally   start with some vector space $\cal X $, 
on the   field  $\batonC$ of complex numbers.

\noi  Then we  extend the  field   of the scalars, by
imbedding the real numbers into a commutative 
 algebra, say  $\batonB$ which,  
as an algebra, is a vector space on the reals but can be complexified in the
usual way.
 By complexification of $\batonB$ we obtain
${\batonB}^\cpx$  equipped with the involution $*$ (we say that  ${\batonB}^\cpx$ is an 
{\sl involutive extension} of  $\batonC$). So
$$  \batonR  \mapsto  \batonB,   \qquad  \quad
{\batonC}   \mapsto     \batonB ^ \cpx                 $$
 The elements of $\batonB$ (resp. $\batonB ^\cpx$)  will be called 
{\sl extra-real} (resp. {\sl extra-complex}).

\noi  This extension entails the possibility of  making linear
combinations of the elements of $\cal X$ by extra-complex numbers,
which uniquely  defines~\cite{bourbex}   a  {\sl module}   over  $\batonB ^\cpx$, 
denoted as  ${\cal X }^\sharp$. 
When ${\batonB}$  actually  is a {\em field}, its complexified  also  is a field,
say  $\batonB ^\cpx =  \batonK$. In this case   ${\cal X }^\sharp$  is a 
{\em vector space} on $\batonK$     and  any Hermitian {\sl map}
$$ \cal X  \times  \cal X  \mapsto  {\batonB ^\cpx}       $$
gets extended as a Hermitian {\sl form}
$$ \cal X ^\sharp \times  \cal X ^\sharp \mapsto  \batonB ^\cpx   $$

\noi We shall carry out this procedure in various spaces of
functions and distributions.  Note that 
$ ( {\cal X}_1  \oplus  {\cal X}_2 ) ^\sharp  =  
    {\cal X}_1  ^\sharp   \oplus  {\cal X}_2  ^\sharp $.   


\medskip
\subsection { Notation}

\noi   ${\calS}$ is the space of test functions (Schwartz space) and 
 ${\calS} '$  is the    space   of tempered distributions. 

\noi $d/dx$ and  $D$ denote  differentiation respectively in the sense of distributions and 
in the sense of functions. 

\noi  The Heaviside step function is noted as  $ \eta (x) $.
Thus
$  \eta (-x) = 1- \eta (x) $.


 \noi
 $\calp$ is the space of polynomials in the
single variable $x$, with complex coefficients.

\noi  $\calq$ is the space spanned by finite complex combinations of the various
 monomial waves    $x^m  e^{iax}$,  where  $m$ is a non-negative integer and  
$a \    \in  \batonR $.  Fixing  $a$  defines  $\calq _a$.

\noi  Let  $\Del$  be the space of  the distrtibutions that have a compact 
discrete support.  
 Let  $\Del _a$ be the  vector space (over  $\batonC$) of the distributions
that have their support concentrated at the point  $x=a$.

\noi  $\calp ^\sharp $ is the space of polynomials in the  single variable
 $x$, with  extra-complex coefficients, similar conventions define 
$\calq ^\sharp$ and   $\Del _a ^\sharp$.

\noi For all $f \in {\calS} '$ we consider the {\sl translations}:
$$ f(x)   \mapsto   f(x+a)$$
and we define also the {\sl phase translations}:
$$      f(x)   \mapsto     e ^{ibx}  f(x)         $$
where $a $ and   $b$  are arbitrary  {\em real} numbers.

\noi Let us also introduce {\sl space reflection}:
$ f(x)  \mapsto  f(-x)  $


 \noi {\sl Fourier  automorphism}

\noi    After  the Fourier transformation
$   f (x)   \mapsto
   F(u) =  ( 2 \pi  ) ^ {- \half }    \
 \int e^{-iux} f(x) \    dx   $,
if we change the name of  the variable  in $F$,  we obtain
  ${ \overline f} = F(x)$,  say for any distribution,
\beq  {\overline  f} =
  {1 \over \sqrt{2 \pi} } \int e^{-ixy} f(y) \    dy             \eeq
Thus   $ \displaystyle  {\overline  { \overline  f} } =  f(-x)$.
   Parity is preserved by  the  Fourier  automorphism:
when  $f(x) = \pm f(-x)$ then we have that
${\overline  f} (x) = \pm   {\overline  f} (-x)  $.

\noi   The well-known properties of the Fourier  transformation
 entail    for all integer $ n \geq 0$,
 \beq  {
\overline {x^n}} = i ^n \sqrt {2 \pi} \    \del ^\ene (x)
                       \label{mapxn}                 \eeq
               More generally
\beq          \overline  { {x^n} \     e^{iax}  }    =
  i ^n  \sqrt {2 \pi} \    \del ^\ene (x -a)           \label{mappolwav} \eeq
which displays the one-to-one map of $\calq$ onto $\Del$, such that
 $\calq _a$ is mapped onto $\Del _a$.


\subsection {Linear operators in  bra-ket  vector spaces}

Consider  a   bra-ket  vector space  $\cal H $ on  a  field  $\batonK$.
Unless otherwize specified, all the operators we consider  acting in $\cal H$
are linear and everywhere defined.

\medskip \noi
If  $\Omega$ is an operator in $\cal H $, we say that $ \Omega ^\dagger $ is a
symmetric of $\Omega $ provided
$ <\Omega  u, v>  =  <u, \Omega ^\dagger  v>  $ for all
 $u,  \  v   \   \in \cal H $. In general, unicity is not garanteed.
Self-symmetric operators are simply called symmetric.

\medskip

\noi
To any  ordered couple $A , B $ where  $A$ and $B$ are in $\cal H $,
we associate the linear operator $ A \times B $ defined as follows:
$$  A \times B  \   w =  <B, w>   A  \qquad     \forall   w  \in \cal H       $$
An operator of this type will be called  a {\sl ket-bra operator} in obvious
reminiscence of the heuristic Dirac's notation  $ |A><B| $.

\noi  It is a simple algebraic exercise to check that
$  B \times A  $ is symmetric to $ A\times B $.

\noi    In particular:

\noi  {\sl any operator of the form  $u \times u$ (projective operator) is self-symmetric}.

\bigskip
\noi
{\sl Idempotent operators}

\noi
Let $\calh$ be a   bra-ket vector space and $R$ a linear operator such that $R^2 = 1$.
For any $f \in \calh $ we have the unique decomposition
$f = f^{\plu }  +  f^{\moi}$
where the {\sl even and odd parts}  of $f$ with respect to $R$,
               are defined as follows
$$  R  f^  {\scriptscriptstyle \pm} = { \pm}  f^{\scriptscriptstyle \pm}$$

\noi
The question arises as to know if  $R$ is symmetric with respect
 to the Hermitian form; it is easy to see that

\beprop
$R$  is symmetric in $\calh$ endowed with $<.,.>$ iff the even
vectors  are orthogonal to the odd ones.
\enprop

\noi Indeed, consider $f, g$ in $\calh$.  Assume for a moment that  $g$ is odd and $f$ is even.
Then we can write
$$  f =  \half  (1+R)  f,     \qquad \qquad     g =  \half (1-R)  g                           $$
thus
$$ <f,g>  =       <    \half  (1+R)  f,        \half (1-R)  g    >
=   {1\over 4}    < f,    (1+R)(1-R)  g >   =  0       $$
Conversely let us now assume that every even vector is orthogonal to every odd one.
Consider any two vectors, say  $f, g$,   in $\calh$, and split them into even/odd parts.
We can write  $Rf =  f^\plu  - f ^ \moi, \qquad    Rg =  g^ \plu  -  g ^\moi$  thus
$$ < Rf , g > =  <f^\plu , g^\plu >  - <f^\moi , g^\moi >  $$
$$ <f, Rg > =  <f^\plu , g^\plu >  - <f^\moi , g^\moi >   $$
But these quantities are equal, thus  $R$ is symmetric.
[]

\noi The above proposition can be applied to space reflection.
Physically it is satisfactory  that space reflection be symmetric, and from
a mathematical point of view it is very natural to  generalize   the odd/even
orthogonality,  from square integrable functions to distributions.

\noi
{\sl In the sequel we shall consider linear spaces made of functions or distributions,
and  $R$ will be the space reflection:}  $R f(x) = f (-x)$.


\section{General Results}

In this Section we recollect several elementary results that can
be easily derived without specifying which commutative field  of
scalars is taken as extension of $\batonC$. We remember that any
nontrivial candidate  for this extension  is necessarily of infinite dimension on
 $\batonC$.

\noi The words "scalar product"  have been loosely employed in the
literature,  including in previous works of the  author. We need
here a more precise  terminology.

\noi
In order to control whether standard results are recovered  as special cases,
it is convenient to set

\noi {\sl   Definition}

\noi
Let ${\cal H}$ be a  space of  functions or distributions; it  is a vector space on $\batonC$
( resp.  on  $\batonK$). 
Assume that  $\calh$  is   endowed with a Hermitian {\em map} (resp. Hermitian {\em form} )
$$  \calh  \times  \calh   \mapsto   \batonK,            $$
 denoted as $ <.,.>$.
 {\em We say that this map (resp. form )  is  {\em admissible} if
$$  <u,v>  =  \int _{- \infty} ^\infty   u^*  v  dx                                   $$
whenever the integrand  in the right-hand side  is  well  defined and the integral  converges  to a   finite complex number}.

 \noi  Admissibility requires that,  in the sense of $<.,.>$                    the odd distributions are orthogonal to the even  ones.


\subsection{Distributions supported by  the origin}

\medskip
\noi
The most elementary distributions which are not trivial ({\em i.e.} not defined by a locally integrable function)
have a discrete compact support.
 In   particular   consider   $\Del _0$.    It is clear that  $\Del _0$ is stable by space reflection and
by  differentiation.

\noi Note that $R$ and  the operator $d/dx$,
 understood {\em in the sense of distributions},
 are  everywhere defined in     $\Del _0$.

\noi Note that, in the Hilbert framework, the skew symmetry of   $d/dx$ 
(understood in the sense of  {\em functions})
was  ensured by the fast decrease (at infinity) of the square integrable functions.

\noi
In   our   context  we turn to distributions and  we observe that (with respect to the behavior at infinity) the elements of  $\Del$  offer an obvious analogy with the   $C^\infty$
 functions  of  fast decrease.
     Thus  it is natural to demand that $d/dx$ (now understood in the sense of distributions)  be
skew-symmetric with respect to  the scalar product.
This property would render the
momentum  operator $ -i d/dx$ symmetric, which is desirable for
the sake of applications to quantum mechanics; we shall  see later
some implications of this requirement. But let us first consider
other operators  everywhere defined in $\Del _0$. Multiplication
by $x$ maps  $\Del _0$ into itself, according to the well-known formula
\beq 
   x   \del^\eme = -m  \del ^{(m-1)}      \label{xdeleme}       \eeq
 Of course we have the commutator  $ [d/dx , x ]= 1$ as usual.

\noi
Let  ${\batonB}$  be some commutative algebra, extension of $\batonR$,
hence   ${\batonB ^C}$  involutive extension of  $\batonC$.
So $   \batonR    \subset   \batonB    ,   \qquad    \batonC      \subset   {\batonB}^C $.
Suppose that  we have a Hermitian   form
   $\   \Del _0 ^\sharp  \times  \Del _0 ^\sharp   \mapsto {\batonB}^C \    $
 denoted as    $<.,.>$,
 such that  $\disp {d \over dx}$  is  skew  symmetric, say
 $$\disp   < {d \phi \over dx} , \psi >  \   =  \    - <\phi , {d \psi \over  dx} > $$


\noi It is convenient to   define $\zeta _k   \in  \batonB^C  $ as follows, for all integer
$k = 0, 1, 2, 3  \cdots   $
 \beq
\zeta_k =  < \del ^{(k)} , \del >       \label{defzeta}    \eeq
Skew symmetry  of       $\disp {d \over dx}$ immediately   entails  that 
$     \zeta ^  *   _ k  =  (-) ^k   \zeta _ k   $.   
It  is easy to verify:

\beprop
          $ \zeta_{2m+1}  =0,\     \forall m \geq 0$   iff
\beq    <\del ^{(2m+1)} , \del ^{(2n)}  >  =0      \label{oddeven}    \eeq
for all integers $m,n \geq 0$.
In other words, $\zeta _{2m + 1} $ vanishes $\forall m$, iff the odd
distributions  concentrated at the origin are orthogonal to the even ones.
 \enprop

\noi  Indeed we note that, owing to the skew symmetry of  $d/dx$
$$        <\del ^{(2m+1)} , \del ^{(2n)}  >  =
         < (d/dx)^{2n} \  \del ^{(2m+1)} , \del  >  =
         < \del ^{(2m+2n +1)} , \del  >  $$
\beq   <\del ^{(2m+1)} , \del ^{(2n)}  >  = \zeta_{2 (m+n) + 1}
                                          \label{zetimpair}                  \eeq
Now assume that  any  odd distribution in $\Del _ 0  ^\sharp$  is orthogonal to the  all
the even ones,  and vice-versa.
Then the left-hand side of (\ref{zetimpair})  vanishes, which means that  also 
   $\zeta_{2 (m+n) + 1}$  vanishes for all   $m,n$.
  But  any $p \geq 0$  is of the form $p=m+n$ for some couple
$m, n$. Thus every     $\zeta_{2 p + 1}$  vanishes. 

Conversely, if we now assume that  $\zeta_k$ vanishes for all odd indices; it follows that
$\del ^{(2m+1)} $ is orthogonal to      $\del ^{(2n)} $. But in  $\Del _0  ^\sharp$
any odd (resp. even)  element  is a  finite  combination  of  distributions  like
$\del ^{(2m+1)} $ (resp.  $\del ^{(2n)} $). Hence the  {\sl odd/even orthogonality}.
[]


\medskip
\noi      Naturally the structure of our Hermitian form 
depends on the detailed shape of the various $\zeta_k$  as elements
of a specified algebra of scalars. However  the following result
is general

\betheo  Let  ${\em  \batonB } ^C$be some commutative algebra, involutive  extension of
$\batonC$. Suppose that  the space   $\Del _0 ^\sharp$
  is  endowed with  a  ${\em  \batonB} ^C$-valued Hermitian  form,  such that
   $R$ is symmetric and  $\disp {d \over dx}$  is
  skew-symmetric.     Then we have that
\beq    <\del ^\eme , \del ^\ene > =
           (-)^m  \     \zeta _{m+n}     =    (-)^n  \     \zeta   _{m+n}        
  \label{LMP}     \eeq
\entheo

\noi   Proof  $ \qquad$
According to Propo.3 the even vectors are orthogonal to the odd ones. On the one hand
  Propo.4  tells that     $\zeta_{2 p + 1}$  vanishes. On the other hand we know that 
 $\zeta ^* _{2k} =  \zeta _{2k} $. Finally    every  $\zeta _p$ is extra-real.

Skew symmetry of $d/dx$ entails that for all integer $p \geq 0$
$$      <\del ^\eme , \del ^\ene > =
 -  <\del ^{(m+1)} , \del ^{(n-1)} > =  +  <\del ^{(m+2)} , \del ^{(n-2)} > =
                                                              \cdots  $$
$$      <\del ^\eme , \del ^\ene > =   \cdots
             (-)^p   \     <\del ^{(m+p)} , \del ^{(n-p)} >          $$
$$      <\del ^\eme , \del ^\ene > =
 (-)^p   \     <\del ^{(m+p)} , \del ^{(n-p)} >               $$
where  $p= 0, \cdots , n$. In particular, when $p=n$ we obtain
$$    <\del ^\eme , \del ^\ene > =  (-)^n   \        \zeta _{m+n}    $$
By  Hermiticity   of the inner product we also have
$  <\del ^\eme , \del ^\ene > =   <\del ^\ene , \del ^\eme >^*  $,
 or (by exchange of  $m$ with $n$) 
$$       <\del ^\ene , \del ^\eme > =   <\del ^\eme , \del ^\ene >^*  =
 (-)^n  \      \zeta ^*  _{m+n}   $$
which is extra-real.  []


\medskip

\noi Remark

\noi Formulae   (\ref{oddeven})(\ref{LMP}) are not sufficient for
the determination of a Hermitian form on $\Del _0$. 
Indeed  all the scalar products  $<\del^\eme , \del >   =   \zeta _m$ 
remain to be defined. To this end, various
inequivalent choices are possible. For instance, eqs (\ref{LMP})
were postulated in \cite{LMP},
 but  with the assumption  that
all the $\zeta_ {2k}$ were independent in $\batonB ^C$.  In the
present  article  this particular assumption is abandoned in
favor of a  more appropriate choice; see  subsection 3.4  below.

\bigskip
\noi In contrast to the above result,   we have this general
impossibility result:
\betheo
If   space reflection is symmetric
whereas $d/dx$ is skew symmetric, then the multiplicative operator
$x$ cannot be symmetric in $\Del _0  ^\sharp$.
\entheo

\noi Proof

\noi Evaluating   $ <x \del , \del ' >$  and   $<\del ,  x \del ' >$ we  
first find  that  symmetry of $x$ would  make  $\zeta _0$ to vanish.    

\noi
 Then consider  any  positive integer  $k$.  From (\ref{xdeleme}) we have on the one hand
$$ < x \del^{(k)} , \del ^{(k+1)} > = < -k \del^{(k-1)} , \del^{(k+1)} >
  =    -k  <\del^{(k-1)} , \del^{(k+1)} > = -k (-)^{k-1}  \zeta _{2k}     $$
But on the other hand
$$ <\del^{(k)} ,  x \del ^{(k+1)} > =
< \del ^{(k)} ,  - (k+1) \del ^{(k)} > =
 - (k+1) < \del^{(k)} , \del ^{(k)} >     $$
and from  (\ref{LMP}) 
 $$ <\del^{(k)} ,  x \del ^{(k+1)} > =
 - (k+1)  (-)^k   <\del ^{(2k)} , \del >  =  - (k+1)  (-)^k  \zeta _{2k}    $$
Symmetry of operator $x$ would imply
$$ -k (-)^{k-1}  \zeta_{2k}   = - (k+1) (-)^k  \zeta_{2k}   $$
$$ ( 2k +1) \zeta _{2k} =  0$$
Hence $\zeta_{2k} = 0$,
 which would make the scalar product to vanish identically.

\medskip
\noi This result tells that, if we  demand  that odd  vectors   are
orthogonal to the  even ones, we cannot endow the space of the distributions supported by the 
 origin with  a scalar product where $x$ and the momentum  operator
$p = -i  d/dx $  are  simultaneously symmetric.  Our choice is definitely
  in favor of the latter,
because  the most  basic ingredient of  energy is its  kinetic   part, which
 involves momentum  (and not  position).

\medskip
\noi  Remark I.   $\qquad$ In \cite{LMP} we  pointed out  a  similar  situation, for a  particular case where
   $\batonC$  was  extended into a  ring of polynomials in  arbitrarily  
 many indeterminates. The Note 11 of that article  contains a guess about more generality.

\noi   Remark II.  $\qquad$ In a recent article, Almeida and Teixeira \cite{alm} have put forward  "position
operators" which  by-pass this impossibility  \cite{alm}. But their operators (defined by
 spectral decomposition) cannot be strictly identified with our
 multiplicative   operator  $x$ (in fact, in
their framework, there are several "position" operators).

\noi    Remark III.   $ \qquad$
 In  spite  of   the above theorem, we note that
multiplication by a smooth function of polynomial growth, say
$V(x)$ defines a symmetric operator, {\em provided that} at the origin
all the derivatives of $V$ vanish.

\section{Ordered Extension of the Scalars}

\noi     So far we have not specified $\batonB$.
  The most simple extension of the scalars consists in replacing $\batonR$
by  a ring of polynomials with real coefficients (resp. $\batonC$, complex
coefficients).


\noi  In a previous work we had considered a ring of polynomials in
{\em infinitely many}  indeterminates~\cite{LMP}. 
   It would be more  reasonable to  employ  the  ring 
  ${\mbox{I\hspace*{-1mm}P}}$   of polynomials  in one {\sl single} 
 real  variable X,  with {\sl real}  coefficients
 (this ring is an algebra over the field of real numbers).
  It is clear that real (resp. complex) numbers can be identified with
 constant polynomials with a real  (resp. complex) value;
that is  $ {\batonR} \subset {\mbox{I\hspace*{-1mm}P}} $  and
  $ {\batonC} \subset \Plex $.

\noi     Still, in a ring of polynomials, the division is generally not
possible. In order to deal with a {\em field}, that is an algebra with
division, we must at least consider {\sl rational functions},
of the form  $\disp P / Q$ where P and Q are polynomials.
Let  $ {\mbox{I\hspace*{-1mm}F}}$   be the field of  {\em real} rational
functions of the   indeterminate  $X$.
For any  element of $\batonF$ we can speak of the  {\em sign at infinity} \cite{bourb}
which allows to define a total (non Archimedean) ordering of $\batonF$.

  Similarly let $\batonF ^C$  be the field of  {\em complex} rational functions of
 the  variable  $X$.       We have
$$ \batonR \subset \batonP  \subset \batonF  ,    \qquad \qquad
     \batonC  \subset  \batonP ^C   \subset  \batonF ^C             $$
                       For all $a \in \batonF ^C$ we
   have $a^*  a   \in  \batonF$   and    $a^* a    >   0$   unless $a=0$.      See Appendix.


\medskip


\subsection{A Positive Scalar  Product}

From now on we assume that the field $\batonB$ is simply
$\batonF$, hence  $\batonK = \batonF ^C$ .

\noi Our main goal remains  the construction of a scalar product for the
 elementary distributions characterized by a discrete finite support.
But for  convenience of the exposition, it is more easy to start with polynomials
$f(x)$. They are trivial as distributions but, through  Fourier  automorphism, they
  happen to be one-to-one connected with the distributions supported at 
the origin~\footnote{It must be clearly understood
 that polynomials in the variable $X$ and  polynomials in the variable $x$
play very different roles: whereas $\batonP$ is just a device for extending the scalars,
$\calp$ is seriously taken as a functional space.}.

\noi
Let us  for a moment     consider  arbitrary   complex  functions of a real variable  $x$.

\noi  When $f(x) $ and $g(x)$ are  $\batonC$-valued square integrable functions, their usual
scalar product
$\disp     \int  f^* g  \  dx  $
  is the limit, for $X \rightarrow \infty$,    of the integral
\beq
I(X, f,g)  =  \int _{-X} ^X   f^* g  \    dx    \label{intX}
\eeq
 When $f(x) $ and $g(x)$ are {\em not} square integrable,
the usual scalar product   may be divergent, or without a limit.

\noi  This  may  occur in particular when $f$ and $g$ are  {\em  polynomials}
in the variable $x$. In this case, the  integral   above
deserves  to  be considered on its own right.
So we define  a  Hermitian map  
$$ {\cal P} \times {\cal P}  \mapsto      \Plex   \subset \batonF ^\cpx  $$
which  depends on $X$  through  the formula  
 \beq    (f, g)  = I(X, f,g)         \label{bracket}           \eeq
Invoking  the  primitive  $\disp  G(x)  =  \int  _0 ^x    f^* g  dx$  we  observe 
  that   $I(X,f,g)$   is an   odd   polynomial  in the variable $X$
( this  would not be the case for  a  more  general  choice of
the functions $f$ and $g$).

In order to  check   admissibility  let us investigate the cases where the usual 
scalar product exists. We obtain

\beprop   Let $f(x),  g(x) $ be complex-valued polynomials. Then  
$\disp     \int _{-\infty}  ^{\infty}   f^* g  \  dx  $
  is finite iff     $ I (X, f, g ) $   vanishes  for  all  $X$,
which occurs  iff   $f^* g $ is an odd polynomial.
\enprop    

Indeed      $\disp  \lim _{X \rightarrow \infty} I (X, f, g )  $ 
 is finite  iff  $I(X) $  is a 
constant.    Being an odd function,   this constant is necessarily zero.        
Conversely, if  $I(X, f, g) $  vanishes  for  all $X$ then  $G$ is even, thus $f^* g$ is odd, which makes its integral
 (from $-\infty $ to  ${\infty}$)     to    to vanish. []


\noi    To summarize:

\noi
 { \em    $(f,g)$ always    coincides with  
        $\disp     \int  f^* g  \  dx  $
 when the latter converges}
 (in this case $I (X,f,g) $ doesnot depend on $X$).


\bigskip

\medskip
\noi
Now extending the scalars from $\batonC$ to $\Flex$ we replace $\calp$
by ${\calp}^\sharp$  but  stick to the definition  (\ref{bracket}).

\noi
For instance any  element of   $\calp   ^    \sharp $ can be written as
$$  f  =   f_0 (X)  +  f_1 (X) \    x  +  \cdots   +   f_D (X) \
  x^D    $$
where     $D$ is integer   and
$f_0  , f_1 ,  \cdots  f_D$    are complex-valued rational fractions
of the only  variable $X$.


\noi  Here equation(\ref{bracket}) defines  a Hermitian form, say
$$ {\cal P}^\sharp  \times {\cal P} ^\sharp  \mapsto      \Flex    $$
Now $f$ and $g$ depend not only on $x$ but also on $X$. They are
 polynomials in $x$ but rational fractions in $X$.  According
to this scheme
 equation  (\ref{intX}) defines a  rational fraction of $X$ which doesnot
 depend on $x$.
We can say that  the scalar product of two polynomials always has a
meaning as an {\sl extra-complex} number, {\em i.e.} an element of
$\Flex$.  Note that $(f,f)$ is always
 in $\mbox{I\hspace*{-1mm}F}$, we say that it is {\sl extra-real}.
Moreover, for $f=g$  we have the following property

\beprop
The scalar product of a polynomial by itself is non-negative in $\em   \batonF $
\enprop
Proof    $ \qquad$ 
\noi The  integral   $I(X)$  in equation(\ref{bracket}) is a rational function of  the
 variable $X$, with no more poles than those of  $f$ and $g$, which  form
 a finite set of points.  This function      $I(X)$ is     defined for all  $X $  larger
than the maximum of these poles,  and  when  $f=g$,  its  {\em  value} is obviously a
 non-negative real number.   Therefore   $I(X)$ is  positive  {\em as a rational function}
in the sense of   the ordering  in  $\batonF$, see Appendix.[]

\betheo
Let $f \in \calp ^\sharp$. Then  $(f,f)$  vanishes iff  $f=0$.
\entheo

\noi Proof

\noi  In the present context, the vanishing of $ I(X, f, f  )$    means  that this
 rational  fraction is zero for all $X$,  and  the vanishing of $f$   means
 that  $f$  is identically zero, for all values of $X$ and $x$.

If we had only      $I(X_0 , f, f ) =0$    for   some    fixed value   $X_0$
of $X$  we could say  that
$  | f (X_0, x ) |  $      vanishes on the interval
$x \in   [-X_0 ,  X_ 0 ] $,  in other words  $f (X,x) $,  function of two
 independent variables,  vanishes on the segment of the $X,x$ plane
defined by the conditions $\    X=X_0 ,  \qquad   x  \in   [-X_0 ,  X_ 0 ] $.

\noi But our assumption is stronger; it  implies that     $I(X , f, f )$
vanishes for every  positive  $X$. Therefore  the function $f(X, x )$ vanishes on
infinitely many segments of the above type, browsing all the domain defined by the conditions   $  -X \leq x \leq  X$             (this domain is limited by
the straight lines      $X=x$  and  $X=-x$).
Since $f(X, x )$  is rational in $X$ and polynomial in $x$,  its vanishing on this domain
 implies that  $f$ is identically zero.[]

\noi {\bf Corollary}

\noi       ${\calp}^\sharp$  {\em endowed with the Hermitian form}
 (\ref{bracket}){\em  is a positive definite inner product space}

\noi  so that $\calp ^\sharp$  is a Hermitian vector space on $\batonF ^C$.

\bigskip

\subsection{Explicit Formulas}

The monomials $x^m$, with  $  m$ a nonnegative integer,   form a  countable  basis
 of ${\cal P}$ (and also  of  $\calp ^\sharp$).      Thus  insofar as ${\cal P}$ only is concerned, 
 it is sufficient to compute the brackets
$(x^m, \   x^n)$  for all integers $m,n \geq 0$ with help of
(\ref{bracket}). One finds, with $m, n  \geq 0$
 \beq (x^m , x^n )
= [1 + (-1)^{m+n} ]  \     {X^{m+n+1} \over m+n+1}
                                     \label{brakpoly}        \eeq
Remark:  this quantity vanishes whenever  $m +n$ is odd.

\medskip

\noi           {\bf Remarks}

\noi
The multiplicative operator  $x$  is everywhere defined in $\calp ^\sharp$  and
 symmetric with respect to the  scalar product     (\ref{bracket}).

\noi  The derivative operator and the multiplication by $x$ are everywhere
defined on $\calp ^\sharp$.
 Moreover  $\calp ^\sharp$    is obviously stable by the translations
$ f (x) \mapsto  f (x+h) $ with real $h$.

\noi  But our  scalar product of polynomials {\em is not invariant}
 under translations.  Similarly, we observe that the derivative operator
 {\em is not} skew-symmetric with
respect to the  scalar product in  $\calp ^\sharp $.

\noi  However the multiplicative operator $x$ is symmetric on
  $\calp ^ \sharp$.


  \subsection{Polynomial Waves}

{\sl Definitions}

\noi   The function $  x^m  e^{iax}$, where $k$ is a nonnegative integer and
$a \in  \batonR$,  will be referred to as

{\sl  the monomial wave of degree $m$ and wave-number $a$}.

\noi  Linear complex combinations  of monomial waves,
with a finite number of terms, but  possibly including different degrees and
 different wave-numbers, will be called {\sl polynomial waves};
they form a  linear space $\calq$.
Such  functions are generally not bounded at infinity  thus, in the position representation of quantum mechanics they are generally
 not eligible as wave functions of physical systems (with  the remarkable
 exception of plane waves).
However polynomial waves are of some interest because, through the   Fourier
transformation, they are in one-to-one correspondance with  the elements of  $\Del$ (resp. $ \Darp$ ).

\noi
Admitting extra-complex combinations, in other words taking the coefficients
in $\batonF ^C$, provides a   vector   space on   the field  $\batonF ^C$, referred to as
$\calq ^\sharp$.

\noi We have these two interesting  special cases:

\noi Ordinary  polynomials correspond to the {\em wave number} zero,

\noi Plane waves correspond to the {\em degree} zero, by finite linear combination
they span a vector space  denoted as  $\calw$.

\noi   The monomial waves with wave number $a$ span a  vector
 space  denoted as  $\calq _a ^\sharp $.
The  elements of  $\calq _a ^\sharp$  are obtained from those of   $\calp  ^\sharp$ by
the phase-translation:   $f  \mapsto   e ^{iax}  f$.
In particular  we have     $ {   \calq _0  ^\sharp  =    \calp   ^\sharp}                $.






\noi  So we  can  define  a  Hermitian form     
on  each   $\calq _a  ^\sharp$   by     imposing
\beq          (x^m   e^{iax}  ,  x^n  e^{iax} ) =
 (x^m ,  x^n )                                            \eeq
 say according to  (\ref{brakpoly})
\beq             (x^m   e^{iax}  ,  x^n  e^{iax} )  \   =
\      [1 + (-1)^{m+n} ]  \     {X^{m+n+1} \over  m+n+1}
                                     \label{brakwave}        \eeq
Observing that
     $\calq_a  ^\sharp      \cap    \calq _b  ^\sharp      =   \{  0   \}$
for   $a \not=  b$ ,  it is  natural to  look after  a Hermitian form
 defined on the direct sum         $ \calq _a  ^\sharp   \oplus   \calq _b  ^\sharp$.

 \noi  By   formula (\ref{brakwave})  the scalar product 
 is first extended, for every  real number $a$,
to the space ${\calq}_a  ^\sharp $ spanned by the functions $x^m  e^{iax}$.

\noi
This requirement alone would not completely define  the form on the whole
$\calq ^\sharp$, but we naturally  also impose  {\sl admissibility}  {\em i.e.} we
also  demand that    $  (x^m   e^{iax}  ,  x^n  e^{ibx} )  $   reduces to
  $\int x^{m+n}  e^{i(b-a)}  dx  $  whenever this integral exists
 as   a  convergent integral  in the  standard framework,
 which occurs for $a \not=  b$, since  in the sense of
 distributions,  setting  $m+n =r, \quad   a-b =c$  we have
$$ \int x^r  e^{-icx}  dx   =   2 \pi  (-i) ^{-r}    \del ^{(r)}  (c)$$
{\em  which is zero}   for nonvanishing $c$.     Therefore formula
     (\ref{brakwave})  is completed by
\beq  (x^m   e^{iax}  ,  x^n  e^{ibx} ) = 0,   \qquad \quad
          \forall     a  \not=  b    \label{brakwavebis}         \eeq
which states that  $ {\calq}_a  ^\sharp $ and  $ {\calq}_b ^\sharp$  are
orthogonal for $a  \not= b$.

\noi
In  view of Proposition 1  it is now clear   that $(f,g)$ is  defined for all $f,g  \in  \calq ^\sharp$.

\beprop
For all $f \in \calq ^\sharp$  the scalar product  $(f,f)$ is positive, unless it is
zero which corresponds to a vanishing $f$.
\enprop

\noi
The proof is straightforward. Indeed  it was proved for
$f \in \calp   ^\sharp$  and is  trivially extended to any $\calq _a  ^\sharp$ by phase
 translation. Then if  $f = f_1  +   \cdots  + f_r$ with $f_k   \in \calq _k   ^\sharp$,
orthogonality of the  various $\calq_a    ^\sharp$ with distinct indices entails
$ (f,f) =  \sum  (f_k , f_k ) $.[]

\noi  This results ensures this

\noi {\bf Corollary}

\noi   {\sl   $\calq  ^\sharp $  endowed with  $(.,.)$  is a (positive) inner product space}.

\bigskip

\noi  It can be read off  (\ref{brakwave})  (\ref{brakwavebis})     that
\beprop
The multiplicative operator $x$ is everywhere defined on $\calq  ^\sharp$ and is
symmetric with respect to the inner  product.
                                     \label{pro:multipx}            \enprop

\noi  Note also that the derivative operator $d/dx$ is  everywhere
defined on $\calq ^\sharp$.
 Moreover  $\calq   ^\sharp$  is   obviously stable by the translations
$ f (x) \mapsto  f (x+h) $ with real $h$.

\noi
 But our    inner product of polynomial waves {\em is not invariant}
 under translations,  although  it  remains invariant  in the particular case of
plane waves,  as can be  read off     (\ref{brakwave})  (\ref{brakwavebis})  by
making   $m=n=0$.

\noi Similarly, we observe that in general the derivative operator $d/dx$
 {\em is not} skew-symmetric with
respect to the  inner  product in  $\calq  ^\sharp$ (for  instance
with $ m=2, n=0$  we find   $ \disp     ({d\over dx}  x^2 , 1   )  =   2 X^3 /3 $
  whereas  $ \disp     (x^2 ,    { d\over dx}  \    1  )  $  vanishes),  although it  remains
 skew-symmetric in the  particular case of plane waves,  say

\beprop
Differentiation maps $\calw$ into itself  and  is  skew-symmetric
in $\calw$.
\enprop

\noi
Indeed 
    we find
$\disp     ( {d\over dx}  e^{iax} ,   e^{ibx})  =  -ia   (e^{iax} , e^{ibx})  $    and
$\disp     (e^{iax} ,    {d\over dx}  e^{ibx} )  =    ib (e^{iax} , e^{ibx})  $.
According to       (\ref{brakwavebis})
    both are   zero if $a \not= b$,  and  if $a=b$ then we get
$\disp     ( {d\over dx}  e^{iax} ,   e^{iax})  =  -ia   (e^{iax} , e^{iax})$    and
$\disp     (e^{iax} ,    {d\over dx}  e^{iax} )  =    ia (e^{iax} , e^{iax})  $,
which proves  the skew-symmetry  of  $d/dx$  with respect to the inner
product of plane waves.[]


\medskip
\noi Remark:  The space of polynomial waves is stable by differentiation,  by
multiplication by $x$,  by { translations}
and by the   { phase-translations}.

\subsection{The space of  distributions with compact discrete support}

\noi    For every $a \in   \batonR$ let  $\Del _a $  be the space of the  distributions
\beq   \phi =       \phi _0  \     \del (x-a)   +
\phi _1  \     \del ' (x-a)  +
\cdots       \phi _r  \      \del ^ {\ere} (x-a)                      \eeq
with constant complex coefficients, where the nonnegative integer $r$ depends on $\phi$.
The {\em  multiplication}  of two such distributions {\em is not}
defined;  
 we wish  however to define a {\em scalar  product}  of them.

\noi
With this goal in mind  we observe that, defining
\beq
\Del  =  \bigoplus _{a   \in  \batonR}     \Del _a
\eeq
hence , by extension of the scalars
\beq
\Del ^\sharp =  \bigoplus _{a   \in  \batonR}     \Del _a  ^\sharp
\eeq
we have
 $ \quad          \calq  \cap   \Del  =  \{ 0 \}    , \qquad   \quad
                      \calq  ^\sharp    \cap   \Del   ^\sharp  =  \{ 0 \}       \quad        $
and   $\Del $ (resp.  $\Del ^ \sharp$)
 is in one-to-one correspondance with $\calq$   (resp.  $   \calq     ^\sharp$).
Namely the Fourier   automorphism  sends   $x^m e^{iax}  \in   \calq$    to
$i^m  \sqrt{2 \pi} \     \del ^m  (x-a) $, formula    (\ref {mappolwav}).
So
$f  \in   \calq  \mapsto {\overline f}  \in  \Del   $  and
$   \calq _a    \mapsto   \Del _a    $.

\bigskip
\noi As well as the monomials waves        span  $\calq$ (resp.  $\calq ^\sharp$),
 the  concentrated  distributions 
$ \disp  \del ^\eme (x-a) $   with arbitrary $a$ in the real line, 
 span $\Del $
 (resp.  $\Del ^\sharp$).  Thus defining a Hermitian map on $\Del $
 (resp. a Hermitian form on     $\Del   ^\sharp$)  will  be   straightforward;
we proceed as follows.

\noi The brackets in $\Del   $ are automatically deduced from the
brackets in $\calq$, since any  distribution    $\phi$  in  $\Del$  is the Fourier
image   of  a   polynomial wave in $\calq$, say    $\phi =   {\overline f}$
and  we  {\em define} 
\beq      <  {\overline f} , {\overline g}   >     =     (f , g )
                                             \label{transstruc}     \eeq

 \noi So doing, we endow each $\Del _a   ^\sharp $ with a structure
isomorphic to that of $\calq_a    ^\sharp$, and $\Del   ^\sharp$ with a structure
isomorphic to that of $\calq  ^\sharp$.

\beprop   In $\Del     ^\sharp$  endowed with $<.,.>$, the scalar product of a distribution
 $\phi$ by itself
 is always  positive  unless it  vanishes,  which  happens only if   $\phi $ is  zero.
\enprop

 \noi   The proof is straightforward
by Fourier duality  since, according to  Proposition 9,  it is true  for  $\calq ^\sharp$.


\medskip

\noi       From Proposition   \ref{pro:multipx} and   well-known properties
of the Fourier transform, we can state

\beprop   The  differentiation    operator   $ \disp   {d \over dx} $
      acting  in   $\Del  ^\sharp $  is  everywhere defined   and  skew-symmetric
with respect to the scalar product    $<. , . >$.    \label{pro:dsurdx}    \enprop

\noi
Accordingly, the momentum  operator   $ \disp     -i \    {d \over dx}  $
 acting  in   $\Del  ^\sharp$  is symmetric.

\noi
We have seen previously that   $ \disp { d \over dx} $
acting in $\calq    ^\sharp$ { \em is not}
 skew-symmetric; similarly, the multiplicative operator $x$ acting in
 $\Del  ^\sharp$  {\em is not}  symmetric.

\bigskip       \noi
We now turn to explicit formulas. Using the linearity of the Fourier transform,
we easily derive from  (\ref{mapxn})
\beq  <\del ^\eme , \del ^\ene > =
   { i^{m-n}   \over  2 \pi }    (x^m , x^n)                    \eeq
  Now  take   (\ref{brakpoly})  into  account. The bracket in the r.h.s. of (\ref{brakpoly})
 vanishes for odd $m+n$ and equals $2$ for for  even $m+n$.  Let us evaluate
    $ <\del ^\eme , \del ^\ene >$ in the latter case,   setting  $m+n=2k$.
We obtain
\beq  <\del ^\eme , \del ^\ene > =       (-)^{k-n}
\     { X^{2k+1}     \over  ( 2k + 1) \pi }      \label{brakdelpair}      \eeq
In particular  we  can write
\beq      <\del ^{(2p)} ,  \del >
            = (-) ^p \    {  X^{2p+1}   \over (2p+1) \  \pi  }
                                                                    \eeq
Let us introduce this notation
\beq     \del ^{\eme} _\zero  =  < \del ^{\eme} , \del >    \label{dezero}  \eeq
        Hence the formulas
\beq    {\del ^{(2p)} }_\zero    =         (-)^{p}
\     { X^{2p+1}     \over  ( 2p + 1) \pi }      \label{deldezer}      \eeq
 in particular
\beq              X = \pi \      \del  _\zero             \eeq
In agreement  with   the convention made in   
   equation  (\ref{defzeta}), 
we shall  equally make use of  the notation
\beq    {  \del ^{(2p)}  } _\zero    =     \zeta _ {2p}    \label{defzetapair}    \eeq
and   we   can  set    $  \zeta _{2p+1}   =   0    $.
With  this convention it  is   not     difficult to compare the present
formalism with an early (and inequivalent)  approach  that we proposed
many years ago~\cite{LMP}. An important difference between the
choices of $\batonB$ in  Ref. \cite{LMP}  and here respectively, is that the choice proposed
in  \cite{LMP}  was an algebra of polynomials depending on {\em infinitely many}  variables.
That  algebra  was  somehow "too large"  for  physical purposes.

\noi
In contrast, in the present work we   employ the rational functions of a  {\em single}
variable $X$ (thus including the polynomials in $X$), which is a field and has a total ordering.


\bigskip
\noi
Let us summarize: we have defined a scalar product on each of the
following spaces $\calp ^\sharp  , \Del _0   ^\sharp,  \Del  ^\sharp ,  \calq   ^\sharp$.
 In  addition  it is obvious  that
$  \calp  ^\sharp   \subset   \calq   ^\sharp$   and
 $\Del _0  ^\sharp    \subset    \Del ^\sharp$ ,
and the scalar product in  $\Del _0  ^\sharp $ is consistent with the
scalar product in $\Del  ^\sharp$.
 Similarly the scalar
product in $\calp ^\sharp$   gets generalized as the scalar product in $\calq  ^\sharp$.

 Moreover    $\calq  ^\sharp  \cap   \Del   ^\sharp =  \{ 0\}  $
 and the Fourier transformation establishes a one-to-one
correspondance between  $\calp  ^\sharp $ and $\Del  ^\sharp$.


\noi In  most  physical applications,  we  shall have to  unify the  inner-product space
$\Del ^\sharp$ with  some suitable space of "ordinary functions", in order to build
at least a  bra-ket vector space.
  As an  illustration of the method we consider  below  the  potential  for   point
 interaction located at the origin.
   In this simple case, the only singularity is  at $x=0$,  and
  it is sufficient
 to  unify  $\Del ^\sharp _0$  with  a suitable space of  functions.

\section{The one-center point interaction in one dimension}

As an  illustration of the method we consider, in one dimension,  the  potential  for   point interaction
located at the origin.   In this simple case, the only singularity is  at $x=0$.
This potential    has a long story \cite{fad}.  Its contribution  is often represented by   the   heuristic expression    $ \alp  \del (x)$
which is  singular in the usual setting.   The most popular  trick consists in  omitting this term ( hence
keeping  the free-particle Schroedinger operator)   whereas  {\em ad hoc}  boundary conditions are imposed
to the wave function.
In contrast,  Albeverio {\em et al}~\cite{alb}    explicitly write down  the potential as  the operator which multiplies  the wave function
by the  characteristic   function of an  infinitesimal  domain around the origin.
Their approach resorts to the most sophisticated  (and powerful)  tools   of  n.s. analysis.

\noi Here also   we  shall  write down  an expression  for the potential, 
 in a much more simple framework  however.


Physical  intuition  suggests   that "nothing special can happen" outside the  origin, 
therefore it is natural to take the view  that  concentrated distributions  like     $\del ^\eme  (x-a)$ 
 where $a \not= 0$   play  no   role  in this problem.
Insofar as  concentrated  distributions  are concerned,   it is therefore sufficient  that  our space of states 
 include    $\Del ^\sharp _0$.  In the same spirit we shall  also include ordinary functions that are 
  smooth  everywhere {\em  except  perhaps} for a possible  discontinuity at  $x=0$.



\noi  In order to be more precise  we  first introduce  left/right  square integrable   functions as follows.
{\em In the standard framework}, we say that a  function   $f$ is  respectively
{\sl right   or   left   square integrable}   when
 $   \int  _0 ^\infty  f^* f  \      dx  $    or
 $   \int  ^0   _   { -\infty }  f^* f  \      dx  $    converges.  Any  square integrable function is
both  right and left square integrable, and {\em vice versa}. 
 But  there are functions that are   only right (resp. only left) square integrable.

\noi
We say that  a  {\em smooth}    function  $f$ belongs  to  $  {\cal  I}^+ $
   ( resp. $ {\cal I}^-  $  )  when  $f$  and
 all its derivatives are     right (resp. left ) square integrable.
Now if  $f_+$    (resp.  $f_- $  )  denotes the restriction of  $f$  at  the interval
$[ - \infty  ,  0 ]$   (resp.  $ [0 , \infty ] $ ) it is clear that
   $f_+ $   belongs to the Hilbert space  $L^2  (  [0 , \infty ] ,  dx ) $       whereas
   $f_- $   belongs to the Hilbert space  $L^2  (  [ - \infty ,   0 ] ] ,  dx ) $.
These Hilbert spaces are respectively endowed with the  complex-valued inner products
$ (.,.)_+$   and  $ (.,.)_- $.


\noi
Finally we introduce $\calj _0$  as the space  of the functions  that  can
 be written as
 \beq
      h(x)  =   f   \eta  +  g  (1-\eta )    \label{ordinaryfon}
 \eeq
with         $f \in  {\cal I}^+ $  and   $   g  \in   \cal I^-  $.
So   $h$   is a function defined   almost everywhere,
 actually everywhere but at the origin \footnote{In this article "almost everywhere"  means
everywhere except at $x=0$.},  and it defines a distribution. 

\noi    Although $h$ is  generally not defined
 at the origin,    we  extend it to this point,  with an {\sl  average value}
$$     h _  0    =   \half   (  \lim  _ {x \rightarrow  0^+ }   h    +     
   \lim  _ {x \rightarrow  0^- }  h    )     $$
                                                            in other words
\beq       h _ 0   =  \half  ( f(0)  +  g(0)   )                        \label{avragzer}    \eeq
so  that     $ h _  0   $   coincides with $h (0)$ in the particuliar case   where
$f=g$.

\noi  From now on, for typographic convenience, we systematically write, respectively  
$f_0 ,   g_0 ,   $,    etc.  for  $f(0),  g (0)  $  etc.    whenever   
    $f(x)   \in    {\cal I} ^+ $   or 
$ g (x) \in   {\cal I } ^- $.  This convention~\footnote{this notation  should not be confused
 with the notation using indices for  the coefficients
in the development of  a  distribution with compact discrete support.}.
is  consistent with  formula    (\ref{avragzer}).

\noi
 Notice that for a given  $h(x)$ in  $\calj _0$  neither $f$  nor  $g $   is
unique; only   their  restrictions  $f_\plu  $  and  $g_\moi$  are unique.
Clearly  $\calj _0$ is endowed with an inner product:
if  $h_1$ and $h_2$  belong to  $\calj _0$,   with  an  obvious notation similar  to that
 of  equation   (\ref{ordinaryfon})    we have
  $$   ( h_1 , h_2 )  =    \int _{- \infty}  ^\infty  h_1 ^*    h_2   \    dx   =
 (f_1 , f_2 )_\plu     +    (g_1 , g_2 )_\moi        $$

\noi
We observe that,  although  $ {\calj  _0}$  does not   coincide  with  $L^2$,  it
includes  $       {\cal S} $    which is dense in     $L^2$.

\medskip
\noi
 {\sl Definition} 

\noi   For any function    $h   \in  \calj _0$  as in  (\ref{ordinaryfon}),
   $\theta   h $  
   is the   {\em jump of    $h$  accross the origin},  say
$  \          \theta   h    =  f_0  - g_0  $

\medskip

\noi     $\calj _0$    is  stable by $D$,  since 
 $\qquad  Dh =   \eta     f'   +  (1-\eta ) g'  $.

\noi  In contrast,   although  $h$  obviously  defines   a distribution, we observe however that its
 distributional     derivative  cannot belong to  $\calj  _0$.  As well known
\beq    h'  =   Dh  +   (\theta    h )    \         \del                    \label{derivjump}             \eeq

\medskip
\beprop
Let  $ \qquad  F = \eta \    u (x)    +  (1- \eta ) \     s (x) ,
   \qquad          G = \eta \    v (x)    +  (1- \eta )  \    t (x), \    $ 
be  two elements of  $\calj _0$.  Then we   have  that
$ (DF , G) +   (F , DG)   =  - u^* _0  v _0    +   s^* _0    t_0  $
\enprop

\noi Indeed we  have
$$DF =  \eta   u'  +  (1-\eta ) s' ,  \qquad  \        DG =  \eta  v'  +    (1-\eta ) t'    $$
$$  \theta F =  u_0  -  s_0 ,      \qquad  \       \theta G  =  v_0  -  t _0     $$
$$  F _ 0   = \half  (        u_0   +  s_0 )    \qquad  \          G _ 0 = \half  (  v_0   +  t_0 )    $$
Let us    compute
   $$                  (DF , G)    =        (u'  ,v ) _\plu    +   (s'   ,t )_\moi     ,  \qquad \
                         (F, DG)      =        (u , v') _\plu    +   (s ,  t'  )_\moi         $$                                           
On the one hand
  $$      (u'  ,v ) _\plu     +      (u , v') _\plu    =         \int  _0 ^\infty   { u' } ^*   v    dx
                            +       \int  _0   ^  {-\infty }    {u } ^*  v'    dx      $$
where $u , v$  are {\em  right} square integrable; integrating by parts yields
       $$      (u'  ,v ) _\plu     +        (u , v') _\plu    =    -  u^* _0    v_0   $$
On the other hand,  in     $ (s'   ,t )_\moi    +     (s ,  t'  )_\moi  $   the functions    $s , t$   are
{\em  left} square integrable; a similar procedure yields
  $$  (s'   ,t )_\moi     +     (s ,  t'  )_\moi         =      s^* _0    t_0                       $$
Adding these two formulas achieves the proof. []

\noi
We shall work  within  the direct sums  $     \Del _0     \oplus  {\calj   _0}  $  and
its  extension 
$        \Del _0 ^\sharp      \oplus   {\calj  _0}^\sharp   $  which  are stable by    $d/dx$.

\noi  The most general element of    $     \Del _0     \oplus  {\calj   _0}  $   takes on the form
\beq  \Phi  =  \phi  +   F ,       \qquad   \phi \in   \Del _0  ,  \quad   F \in  \calj _ 0   \label{Phigene}   \eeq
In the sense of  distributions,  all the derivatives
$   \disp     F^\ene  =   {d ^n     F  \over   dx ^n}   $
    are in   $\Del _0   \oplus  \calj_0$,  thus all the derivatives
$$ \Phi ^\ene       =     \phi ^\ene  +  F^\ene           $$
 also belong to  $  \Del _0   \oplus   \calj _0$.

\noi  {\sl Definition}

\noi {\em Let      $ \Phi  \in     \Del _0   \oplus   \calj _0$.
     The {\em   pseudo-value    of   $\Phi$    at the origin} is the quantity
\beq  \Phi _\avragzer   =   F_0  +   <\del  , \phi >   \label{pseudoval}                \eeq
where $F_0$  is the average  value of $F$ at  the origin,
 like in formula  (\ref{avragzer})  }.

\noi This definition is a  generalization of the convention made in (\ref{dezero}).

\bigskip 
\noi
{\em By extension of the scalars, Proposition 12  and the definition of average value  carry over to $\calj _0 ^\sharp$ ; 
the definition of pseudo-value carries over  to 
$  (  \Del _0   \oplus   \calj _0 ) ^\sharp $ }.
 
\medskip
\noi  We  want  to  define a  Hermitian {\em form}   on
   $    (   \Del _0       \oplus   {\calj  _0} )  ^\sharp         $;
to this end  it will be sufficient to   define  first  a Hermitian {\em map}
 $  (\Del _0       \oplus   {\calj _0})  \times 
 (\Del _0       \oplus   {\calj _0})   \          \mapsto    \batonF ^C  $.

\noi
So let
\beq     F =   \eta  \       u (x)        +     (1-\eta ) \    s (x)       ,
    \qquad   \quad
        G =    \eta \      v (x)   +      (1-\eta ) \     t(x)            \label{FG}    \eeq
    be two arbitrary  elements of    $  \calj _0$.
For the sake of admissibility  we  impose that   $<F,G> $  is simply
 $\disp   \int _{-\infty}  ^  {+ \infty}  F^* G  \    dx  $.

\noi  
According to Proposition 1,  since  the various $\del ^\eme $   are  a basis  of   $\Del _0 $,   
   we must  give a meaning to     all the  brackets
$   < \del ^\eme   ,  G >   $   and      $   < F ,  \del ^\eme   >   $  where  $F$ or $G   \in   \calj _0$.

 \medskip


\noi  Consider $\Phi , \Psi $ as in  (\ref{Phigene}), say  
$ \Psi = \psi  +  G$.
  Since
$<\del ^\eme , \del ^\ene >  $  was   already   defined, inducing the expression for    $ <\phi , \psi >$,
  the only thing which remains to be done consists in   defining  
$<\phi , G > $ and  $<F, \psi >$.

\noi  By analogy with the case of smooth functions, we postulate

\beq
<\del ^\eme , G >  =  (-)^m  \    (G ^\eme )_ \avragzer
                                      \qquad  \       m  \geq  0          \label{delemPsi}                                 \eeq
In order  to  get  a  {\em Hermitian}   map  we are obliged to   complete by
the   conjugate   formula, say
\beq   <F , \del ^ \ene > =      (-)^n  \    ( {F  ^* }^\ene )_ \avragzer
              \qquad  \       m  \geq  0                              \label{Phidelen}                                  \eeq
Then,  since the distributions   $\del ^\eme $  are a basis of  $\Del _0$,  the   above  formulas    automatically
define  the  sesquilinear   maps
$$    \Del _0 \times    \calj_0     \mapsto    \batonF ^C  ,
\qquad  \
             \calj_0  \times  \Del _0       \mapsto   \batonF^ C      $$
hence     (with  help of Proposition 1)                        
 the   Hermitian   {\em map}   
  $ ( \Del _0  \oplus  \calj _ 0 )  \times     ( \Del _0  \oplus  \calj _ 0 )
      \mapsto    \batonF ^C                   $.   Finally  {\em extending the algebra of the scalars}  provides   the Hermitian {\em form } on   
  $( \Del _0  \oplus  \calj _ 0 )^\sharp$. For instance 
(\ref{delemPsi}) and  (\ref{Phidelen})  keep  being  valid when  $F$ and $G$  belong to  $\calj _0 ^\sharp$,  etc.

\noi  The form is admissible:    we   recover the inner product  in 
$\Del _0 ^\sharp $  as a particular case, and  also
the ordinary  scalar product  $ (\del ^\eme , G )   $ in the simple case where   $G$  is an  everywhere defined  and smooth  square integrable function.

\bigskip

\noi
Let us now  exhibit a nice property of    $d /dx$  in the bra-ket space
$    (\Del _0  \oplus  \calj _ 0  ) ^\sharp  $. It is convenient to observe first the following
 \beprop
For  $\phi  \in  \Del _0 ^ \sharp $   and   $G   \in   \calj_0  ^ \sharp $  we  have
\beq     <\phi , G ' >  +  <\phi ' , G >  =  0  \label{proplem}   \eeq
\enprop

\medskip
\noi         Proof:
On the one hand   writting  formula (\ref{delemPsi})  for  $m+1$    yields
\beq       <\del ^ {(m+1)}  ,   G >  =   (-)^ {m+1} \     G  ^{m+1}  _\zero
                                                                                                     \eeq
On the other hand, writting    (\ref{delemPsi})   for  $G ' $    yields
\beq
  <\del ^ \eme  ,   G   '  >    =      (-)^ m  \             G  ^{m+1}  _\zero
\eeq
Say  $  \disp \     \phi  =  \sum  \phi _m  \del ^\eme \    $  hence
         $  \disp   \phi  '   =  \sum  \phi _m  \del ^  {(m+1)}    \      $
we get
$$  <\phi , G ' >     =    \sum            (-)^ m      \phi _m        G  ^{m+1}  _\zero     $$
$$     <\phi  '   , G  >     =     \sum            (-)^  {m +1}    \phi _m      G  ^{m+1}  _\zero         $$
but    $  (-)^ m $   and  $        (-)^  {m +1}$   are opposite.[]

\medskip
\noi  We are in a position to derive this  result


  \betheo
  Defining the Hermitian  form  with help of the formulas
(\ref{delemPsi}) (\ref{Phidelen})      ensures  that  $ d/dx$  is skew-symmetric    in
     $ (   \Del  _ 0            \oplus           \calj _0   ) ^\sharp $
 \entheo

\noi  Proof

\noi
Consider  in   $    (   \Del _0       \oplus   {\calj  _0} )  ^\sharp                 $  two  elements
$\     \Phi  = \phi  +  F, \qquad          \Psi   =   \psi +    G     \     $  where  $\phi$ and  $\psi$
  are in $\Del _0 ^\sharp$  and  $F, G$ as    in     (\ref{FG}),
with   $u , v   \in  {\cal I}^{+ \sharp}$  
 and   $s , t  \in  {\cal I} ^{- \sharp}$.                           
Compute   $ <\Phi '  , \Psi >   +       <\Phi    ,  \Psi  '   >  $.  We get
$$  <\Phi '  , \Psi >   =     <\phi ' , \psi >    +   <\phi  '  , G >  +  <F '  , \psi >  +  < F'  , G >$$
$$  <\Phi    ,  \Psi  '   >  =   <\phi  , \psi  '  >    +   <\phi    ,  G'     >      +        <F  , \psi  '   >      +  < F  , G '  > $$
Taking into account the skew-symmetry of $d/dx$  in     $\Del _0 ^\sharp$   yields  a  first cancellation,
we are left with
$$     <\Phi '  , \Psi >   +       <\Phi    ,  \Psi  '   >      \             =     $$
\beq    <\phi  '  , G >  +  <F '  , \Psi >  +  < F'  , G >         +   <\phi    ,  G'     >      +        <F  , \Psi  '   >      +  < F  , G '  >
                                                                                              \label{rouge}        \eeq
\medskip
\noi  Now  let us
  calculate   the contribution of   $< F'  , G >    +  < F  , G '  >$.
According to  the rule (\ref{derivjump})  we find
  $$     < F'  , G >    +  < F  , G '  >   =
< DF  +  \theta  F  \      \del  ,   \         G>  +   
<F, \     DG  +  \theta  G  \         \del  >     $$
\beq      < F'  , G >    +  < F  , G '  >   =
<DF , G>   +  <F,  DG>   +    (\theta F )^*    \     G_0   +   \theta  G  \       F^*  _0                                                            \label{VI}                 \eeq
But          $\    \theta F =  u_0  -  s_0 ,   \qquad     \
                \theta   G  =   v_0  - t_0   ,    \            $                 hence with help of     Proposition 12
     $$     < F'  , G >    +  < F  , G '  >   =    -u^* _0     v_0   +  s^* _0   t_0
+ ( u_0 ^*  - s_0  ^* )   G_\zero    +  (v_0 - t_0 )  F^*  _\zero                 $$
Remember that
$\disp   \qquad            F_0 = \half  (u_0 +  s_0 ),   \qquad
  G_0 =  \half  (v_0 +  t_0 )    \qquad              $  thus
   $$     < F'  , G >    +  < F  , G '  >   =   $$
$$  -u^* _0     v_0     +  s^* _0   t_0
+   \half    ( u_0 ^*  - s_0  ^* )    (v_0 +  t_0 )
+ \half    (v_0 - t_0 )   (u^* _0 +  s^ * _0 )       $$
which identically vanishes, so
\beq     < F'  , G >    +  < F  , G '  >   = 0    \label{VII}               \eeq
We remain with
\beq   <\Phi ' , \Psi >       +       <\Phi    ,  \Psi  '   >       =
    <\phi ' , G>  +   <\phi ,  G' >  +   <F' , \psi > +  <F , \psi '  >    \eeq
which vanishes in view of   Proposition 13.[]


\subsection{Solving the  Schroedinger  equation.}

\bigskip
\noi We are now in a position to consider the Schroedinger  equation with a
simple example of   point  interaction.

\noi
In view of the considerations made in subsection   1.3
 let us first write down the
"singular"  potential  in terms of the ket-bra  operator
 $\del \times \del$.  We limit ourselves to the case of an attractive potential  so we write the wave equation as
\beq   y''   +   \lam  y     =   \alpha    (\del \times \del ) y  
 \label{eqonde}   \eeq
where  $\alp$ is a  {\em negative  real}  number,  say   $\alp  =  -2 \beta$.
Having in mind  physical  applications  we look  for  a  real   standard 
eigenvalue  $\lam $.
  Here $y$ is supposed to be an element of
  $ { \calj} _0 ^\sharp   \oplus  \Del _0 ^\sharp $ and the derivatives are meant
in the sense of distributions.


\noi    Since we seak solutions in     $ (\calj _0  \oplus  \Del _0 )^\sharp $ we  write

\beq   y = f \   \eta + (1 - \eta ) \    g  + \phi            \label{y}          \eeq
where
 $f  \in   {\cal I} ^{ +  \sharp} ,     \qquad       g   \in   {\cal I}  ^   {- \sharp} ,
   \qquad   \phi \in \Del _0 ^\sharp $.

\noi  Since $\   \eta ' = \del ,   \quad   (1 - \eta ) ' = - \del \     $, we obtain
$$ y'' =   f'' \eta +  g'' (1 - \eta ) +  2 (f' - g' ) \del
+ (f-g) \del    +   \phi '    $$
But  $ h \del ' = h_0  \del '  -  h' _0  \del  $   for  all  function  $h$.
Hence 
\beq   y '' =    f'' \eta +  g'' (1 - \eta ) + (f'_0 - g'_0 ) \del
+  (f_0 - g_0 ) \del '    +  \phi ''          \label {y''} \eeq
\beq    y'' + \lam y =
(f'' + \lam f )\eta + (g'' + \lam g ) (1 - \eta )        + (f'_0 - g'_0 ) \del
+  (f_0 - g_0 ) \del '    + \phi ''  + \lam \phi    \label{y''bis}  \eeq
Let us now evaluate the contribution  of the potential,
$$ \alp   \   (\del \times \del ) y = \alp  <\del, y> \del                 $$
$$ <\del , y > =  \half f_0 + \half g_0    + <\del , \phi>         $$
\beq          \alp (\del \times \del ) y =
{\alp  \over 2 }  ( f_0 + g_0 ) \del    +   \alp    <\del , \phi >  \del
                                                                      \label{poty} \eeq
Note that  (\ref{y''})(\ref{y''bis})   and   (\ref{poty})  hold  true
 for all $y$
 of the form (\ref{y}) irrespective of the Schroedinger equation.
But now  insert
     (\ref{y''bis})    and (\ref{poty}) into  (\ref{eqonde});  we identify
and get    separately
\beq    (f'' + \lam f )\eta    +    (g'' + \lam g ) (1 - \eta )  = 0
                                                                 \label{spliteq}           \eeq
\beq   {\alp \over 2}   (f_0 + g_0)\   \del   +     \alp  <\del , \phi >  \   \del   =
(f'_0 - g' _0) \   \del  + (f_0 - g_0) \   \del'
+ \phi  ''  +  \lam \phi                             \label{identf}      \eeq
{\sl Remark}   

\noi Since  $\phi $ is a  {\em finite} sum,   we can write
\beq   \phi = \sum _0 ^r  \phi _n \del ^ \ene        \label{phi}             \eeq
and  it can be read off  (\ref{identf}) that  the development  of  the
distribution $\phi '' + \lam  \phi$  cannot  involve  derivatives of  $\del$
 higher than   $\del '$.

\beprop   In     (\ref{y})  necessarily  $\phi $ vanishes              \enprop

\noi The proof  is {\em ab  absurdo}.
    Assume that  $\phi$ is actually of order $r$, in other words
  $ \phi _r$   doesnot vanish.  Differentiating    (\ref{phi})  we obtain
$ \disp     \phi '' =    \sum _0 ^r  \phi _n \del ^ { ( n  + 2) }       $,
 hence  the  development
$$  \phi '' + \lam  \phi =
                            \lam \phi _0 \del + \lam \phi _1 \del '      
 + \lam \phi _2 \del ''  + \cdots   + \lam \phi _r \del ^{(r)}        $$
$$ +  \phi _0 \del '' + \cdots  + \phi _{r-2} \del ^{(r)}
+ \phi _{(r-1)}     \del ^{(r+1)}   +   \phi _r \del ^{(r+2)}              $$


\noi

In view of  the  Remark  above, it is clear
 that   $\phi _r $ must vanish,  contrary to our assumption.[]

\medskip
\noi On the one hand  (\ref{spliteq}) splits into these two statements
\beq   (f''_\plu  +  \lam  f _\plu   ) =     0,   \qquad
       (g''_\moi       + \lam g_\moi    )  = 0             \label{sturmliou}  \eeq
\noi  On the other hand, in view of  the above Proposition  we can
rewrite    (\ref{identf}) simply as
 \beq           {\alp \over 2}     (f_0 +  g_0)\ \del       =
   (f'_0 - g' _0) \   \del  + (f_0 - g_0) \    \del '
                             \label{boundry}      \eeq
which provides us with  two   boundary conditions
\beq      { \alp  \over 2}   (f_0 + g_0)    =    (f'_0 - g' _0)         \label{58'}     \eeq
\beq      f_0 = g_0     \label{58"}    \eeq

\noi    Up to arbitrary and irrelevant (smooth)  modifications
 of  $f_ \moi$    and   $g_\plu$,
the most general solution  to   (\ref{sturmliou})  is,
  if $\lam \geq 0$
$$ f = A^\plu  \exp (i \sqrt{ \lam} x) +    A^\moi   \exp (-i \sqrt {\lam} x)     $$
$$ g = B^\plu    \exp (i \sqrt{\lam} x) +    B^\moi   \exp (-i \sqrt{\lam} x)     $$
and  if   $\lam <0$,    setting  $\lam = -\rho$,
$$ f = a^\plu   \exp ( \sqrt {\rho} x) +    a^ \moi   \exp (- \sqrt {\rho} x)     $$
$$ g = b^\plu   \exp ( \sqrt {\rho} x) +    b^\moi   \exp (- \sqrt {\rho} x)     $$
where  $A^\pm , B^\pm , a^ \pm , b^\pm$ are in $\batonF$.
 But $f$ (resp. $g$) is supposed to be right (resp. left ) square integrable, thus
{\em  within   $ { \calj}_0 ^\sharp   \oplus  \Del _0 ^\sharp $    no eigenstate
  corresponds to   $\lam    \geq  0$.}

\noi  Let us  turn to the possibility of a negative $\lam$.
Since $f$ (resp. $g$)
is supposed to be right (resp. left ) square integrable,  it is clear that
$a^+$  and  $b^-$ must vanish.
Therefore
\beq
f=  a^-  {\rm exp}   ({ -  {\sqrt \rho} \    x}) ,  \qquad \quad
g= b^+  {\rm exp}   ({ {\sqrt \rho} \    x} )
                      \label{61}                                    \eeq
Hence
$$ f_0 = a ^- ,  \qquad  g_0 = b^+ ,  \qquad
f' _0  = - a^-    {\sqrt \rho}      , \qquad    g'_0  =  b^+    {\sqrt \rho}   $$
Now condition (\ref {58"}) yields  $a^- = b^+$ and  (discarding the trivial solution $y=0$)
condition     (\ref {58'})  yields       $\alp  =  -2  {\sqrt \rho }$.
 Thus the only possibility is
\beq   y =  {\rm const.}  \    ( {\rm e}  ^{- \beta  x} \     \eta (x)     +
          {\rm e}  ^{ \beta  x} \     \eta (-x)       )    \label{62}                   \eeq
The constant factor is arbitrary, but  standard solutions  correspond to choosing
this factor in $\batonC$.

\medskip
\noi     Conversely, it is  easy to check that (\ref{62}) actually satisfies (\ref{eqonde})  provided 
$  \lam = - \beta ^2 = - {\alp ^2  \over 4}           $.
This calculation uses the formula
     $$ <\del ,  \    {\rm e}  ^{- \beta  x} \     \eta (x) > =
        <\del ,  \    {\rm e}  ^{ \beta  x} \      \eta (-x) >  =          \half   $$
obtained  from  (\ref{avragzer}).  []
\noi   We summarize:

\betheo
In  $ ( \calj _0  \oplus  \Del _0  ) ^\sharp $,  the linear operator 
$ \disp  - {d^2  \over  dx^2} +  \alp  (\del \times \del )  $, with real negative $\alp =  -2 \beta $,
  has  the  eigenvalue 
$\disp  - {1 \over 4} \alp ^2$ and the  eigenvector given  by  (\ref{62}).
\entheo

 \section{Conclusion and outlook}

\noi
In  this work a   positive definite Hermitian form  was constructed  for the
space of  distributions  with compact   discrete support.   The values of this form  belong to
a field  extension of   $\batonC$    which,   in  spite  of  its  simplicity,  allows  for
considering   some  infinitesimals  and   some   infinitely large  numbers.
This extension is in an  obvious  sense minimal,  and  further extensions are certainly
desirable. An  amusing  open question is how far can we go on  doing  n.s. {\em calculus}
 with elementary methods that ignore n.s. {\em analysis}.

\noi
Let us emphasize that here  the distributions are taken for what they  usually  are in 
the environment of   standard  quantum 
 mechanics,  and {\em not   replaced}  by  functions
 $ {   \  }^*  \batonR     \mapsto    {   \  }^*  \batonR $   with  infinitesimal support~\cite{Li}.
In this matter we  remain rather conservative: the  distributions  we  consider  can be   seen  as  $\batonF  ^C$-valued  linear functionals,    but their argument $x$  runs inside     the {\em standard}  real line $\batonR$.
It would be interesting to make a contact with some results of reference~\cite{Li} but this difference
 in the status of the distributions renders such a task difficult.

The structure of  bra-ket  vector space  seems to be the  most general  framework 
 available    for implementing  the  machinery  of  Dirac's  formalism  in  a  flexible  but rigorous way;  
 in  most   physical situations  a  richer structure   would  be desirable,  at  least  that of   inner-product  space, thus more  work is needed in this direction.

For elementary applications to the Schroedinger equation,   the  differential operator $ d/dx$  must  be  understood in the sense of  
distribution theory   and  some subspace of  $\Del ^\sharp$  is to be    unified  with  a suitable  space  of  functions,
both spaces  getting   imbedded  into  a   larger  vector space  equipped   with   a    Hermitian  form.  
Further investigation  is needed in order to determine  the  cases  where  this form   remains positive definite.  

\noi Naturally this  line of research has  obvious limitations:  
we  avoided   any   kind of topological  considerations,    preferring   to focus  on  the  direct  computational  
mechanism   provided   by   the   bra-ket    framework.

\noi However we expect  that in several  cases  this approach   will      legitimate  a lot of heuristic  calculations that would  seem meaningless otherwise.

\noi
 As  a  very  simple  example we  considered the  point interaction in one dimension. The potential was treated 
in a   natural  and  intuitive manner,    being symmetric  and everywhere   defined as a linear operator. Playing with cancellations of infinite quantities,  we  re-derived  the usual eigenvalue and eigenfunction corresponding to the bound state. 
Naturally we  remain aware of the possibility  to attain the same result through  the  sophisticated methods of   n.s. analysis displayed in \cite{alb}, but  one of our  goals  was precisely to  provide an elementary  formulation available  to  every  theoretical  physicist.


\medskip

\section{  APPENDIX}

\noi  {\bf  Minimal extension of the real numbers}.

\noi
 We take  $\batonB = \batonF$  where $ \batonF$    is
the field of real rational functions in one real variable $X$
(rational functions being characterized by
their  form  $f = P/Q$ where $P$ and $Q$ are polynomials).

\noi 
For large enough  $X$    any  nonvanishing
  rational   function    $f$ takes on a definite sign, so that
speaking of the {\sl sign of $f$ toward infinity} makes sense  . We
say that $f$ is positive or negative (toward infinity) according
to this sign.   Moreover this sign is compatible  
with the    multiplication in $\batonF$,  that is
$ {\rm sign} (f) {\rm sign} (g) = {\rm sign} (fg) $,
     which permits to define a  {\em  total   ordering}
 among all the rational functions,  just  by   saying that    $ f > g $ iff
$f-g$ is positive  toward    infinity.

\medskip

\noi  Most usual properties of the real numbers carry over to the
field of rational fractions. For instance: 
     $ f> 0$ and $g> 0$ imply $f+g >0$.

\noi When $f > 0$, having $g < h$ implies $fg  < fh$. When both
$f, g > 0$, then also $fg$ is positive; moreover having that $f>g$
implies that $f^2> g^2$.

\noi  Then it is possible to define an absolute value:  $||f|| =
f$ if $f \geq 0$ and $||f|| = -f $ otherwise. It is clear that
$||f||$ is a positive extra-real unless $f=0$,  and that  every
square of a non-vanishing rational fraction is positive.

 \noi  Finally we can check that for
all couple $f,g$ we have
 $ ||a+b|| \leq ||a|| + ||b $.

\noi Any  rational fraction  $f(X)$ has a {\sl degree}
\footnote{When $ P, Q$ are polynomials, $ \deg P/Q = \deg P - \deg
Q$ } noted as  $\deg f$, which is non-negative for polynomials. We
have these inequalities

 $$\degr (a+b) \leq  {\rm max} (\degr   \    a ,  \degr \    b )$$
 $$  \degr (ab)  =  \degr \     a  +  \degr \     b                    $$
 obvious for polynomials, then  easily extended to rational functions.

  \noi As well known, every $f \in \batonF$  admits a unique decomposition
$     f =   P  +  \epsilon    $
where  $P (X) $ is a polynomial and $\epsilon  (X) $ is a rational function 
of  negative  degree.

\noi  According to  the ordering   we  have   $\epsilon  (X)  <  c      <   X^p   $  
       whenever $p$ is a positive integer and $c$  is a   positive  constant,   
so   we can interprete  the positive   powers of  $X$   as   {\sl "infinitely
 large"} numbers, and  $\epsilon$ as {\sl "infinitesimal"}.
 
\noi  
Let  $a$ be the constant term in $P(X)$. Then $a + \epsilon $ can e called the  {\sl finite part} of $f$,
  and  $a$ its {\sl standard part}.
This possibility of extracting a "standard part" seems to be a    particular feature of $\batonF$.
 It would probably not survive in  further extensions  from    $\batonF$
to a larger   ordered  field.

\bigskip

\noi Although the present scheme has been introduced by elementary
methods~\cite{bourb}, it  mimicks the behavior  of the  "hyperreal  numbers"
considered in the framework of n.s. analysis. The reader
who is familiar with n.s. analysis  can understand this as
follows:

\noi Let  $  \kappa   \in \     ^* \batonR $ denote any  infinitely
large and positive hyperreal.   Then  the structure of $\batonF$
is isomorphic to the class formed  by all the hyperreal numbers
that can be written as a rational function of $\kappa$ (this class
is an algebra on  $\batonR$).

\noi   Due to  the arbitrariness in the
choice of $\kappa$ we conclude that $\batonF$ can be imbedded into $
^* \batonR $  {\sl in infinitely many different ways}.
But there is no preferred correspondance. Note that, in contrast to
 n.s. analysis, our approach  has  no concept of "infinite integer".

   \end{document}